\documentclass[12pt]{iopart}
\usepackage{graphicx}
\usepackage{citesort}
\usepackage{indentfirst}
\usepackage[justification=justified,singlelinecheck=true]{caption}
\usepackage{iopams}
\expandafter\let\csname equation*\endcsname\relax
\expandafter\let\csname endequation*\endcsname\relax
\usepackage{amsmath}

\begin{document}

\title{Low-field magnetotransport in graphene cavity devices}

    \author {G. Q. Zhang$^1$ N. Kang$^1$ J. Y. Li$^{1,2}$ Li Lin$^3$ Hailin Peng$^3$ Zhongfan Liu$^3$ and H. Q. Xu$^{1,4}$}
    \address{$^1$ Beijing Key Laboratory of Quantum Devices, Key Laboratory for the Physics and Chemistry of Nanodevices and Department of Electronics, Peking University, Beijing 100871, P. R. China}
    \address{$^2$ Academy for Advanced Interdisciplinary Studies, Peking University, Beijing 100871, P. R. China}
    \address{$^3$ Center for Nanochemistry, Beijing Science and Engineering Center for Nanocarbons, Beijing National Laboratory for Molecular Sciences, College of Chemistry and Molecular Engineering, Peking University, Beijing 100871, P. R. China}
		\address{$^4$ Division of Solid State Physics, Lund University, Box 118, S-221 00 Lund, Sweden}

    {\bf \ead{\mailto{nkang@pku.edu.cn},\mailto{hqxu@pku.edu.cn}} }

    \begin{footnotesize}
    {
    \hspace*{3.6pc}
    }
    \end{footnotesize}

\begin{abstract}
Confinement and edge structures are known to play significant roles in electronic and transport properties of two-dimensional materials. Here, we report on low-temperature magnetotransport measurements of lithographically patterned graphene cavity nanodevices. It is found that the evolution of the low-field
magnetoconductance characteristics with varying carrier density exhibits different behaviors in graphene cavity and bulk graphene devices. In the graphene cavity devices,
we have observed that intravalley scattering becomes dominant as the Fermi level gets close to the Dirac point. We associate this enhanced intravalley scattering
to the effect of charge inhomogeneities and edge disorder in the confined graphene nanostructures. We have also observed that the dephasing rate of carriers in the cavity
devices follows a parabolic temperature dependence, indicating that the direct Coulomb interaction scattering mechanism governs the dephasing at low temperatures. Our
results demonstrate the importance of confinement in carrier transport in graphene nanostructure devices.
\end{abstract}

\maketitle
\section*{Introduction}
Recently, study of graphene-based nanostructures has attracted increased attention due to their exceptional electrical, optical, and mechanical properties and
their potential applications in carbon-based nanoelectronics and
optoelectronics \cite{ChenRosenblattBolotinEtAl2009,GuettingerFreyStampferEtAl2010,HanOezyilmazZhangEtAl2007,YangCohenLouie2007}. Several graphene nanostructures,
such as nanoribbons, grain boundaries, and etched nanostrcutures, have been fabricated and used to explore the properties of quantum-confined electronic states in
graphene \cite{RitterLyding2009,LiWangZhangEtAl2008,MolitorJacobsenStampferEtAl2009}. Because of a small device dimension, the electrical transport properties of
graphene nanostructures are expected to be sensitive to scattering from disorder potentials and graphene
edges \cite{ZhangBrarGiritEtAl2009,GuettingerStampferLibischEtAl2009,SongLiYouEtAl2015}. The studies of the effects of confinement and interference, and of their consequences on
carriers transport in graphene nanostructures are important not only in understanding of the fundamental low-dimensional quantum transport properties of Dirac electrons but also in advances of graphene-based nanodevice applications.

The quantum interference corrections to the conductivity of graphene are different from the conventional two-dimensional (2D) systems due to the chirality of the Dirac fermions and the induced Berry phase $\pi$ of transport carriers in graphene \cite{KotovUchoaPereiraEtAl2012,SarmaAdamHwangEtAl2011}. In conventional 2D systems, in the presence of the time reversal symmetry, the phase shifts of carriers propagating along a closed trajectory in clockwise and anticlockwise directions are the same, which causes constructive interference and leads to the so-called weak localization (WL).
The situation is different in graphene--due to the presence of the Berry phase $\pi$, the interference becomes destructive, leading to weak antilocalization (WAL) of charge carriers in graphene \cite{AndoNakanishiSaito1998}. The carriers in graphene are chiral, that is, they have a pseudospin which is always parallel to the momentum in one valley while in the other one is antiparallel. An electron (or a hole), scattered by disorder, could not be scattered back within the same valley because the pseudospin can not be flipped and the chirality should be conserved. The theory of quantum interference in graphene\cite{McCannKechedzhiFal'koEtAl2006} shows that in addition to the dephasing rate ${\tau_\varphi}^{-1}$ caused by inelastic scattering, there are other elastic scattering contributions: intervalley scattering rate ${\tau_i}^{-1}$ caused by short-range potentials such vancacies and sample edges \cite{Peres2010,TikhonenkoKozikovSavchenkoEtAl2009}; intravalley scattering rate ${\tau_s}^{-1}$ caused by long range potentials such as ripples, dislocations, and charged scatters; and intravalley   scattering rate ${\tau_w}^{-1}$ caused by trigonal warping \cite{AndoNakanishiSaito1998}. The trigonal warping effect destroys the inversion symmetry of the energy band around the Dirac point within the same valley, leading to nonvanishing back scattering. Specifically, the equi-energy lines around the Dirac point are no longer circular contours, but deform into triangular contours, because of characteristic trigonal warping effect with increasing energy in graphene. Despite intensive studies of the WL and WAL effects in quantum transport of bulk graphene flakes \cite{MorozovNovoselovKatsnelsonEtAl2006,Fal'koKechedzhiMcCannEtAl2007,TikhonenkoHorsellGorbachevEtAl2008,WuLiSongEtAl2007}, few work has studied the effects of confinement on these scattering mechanisms in graphene nanostructures.

In this work, we study the role of confinement in the magnetotransport properties of a graphene cavity at the mesoscale.
We analyze the quantum interference effects in both graphene bulk and cavity devices. The characteristic scattering rates have been determined from the analysis of the magnetoconductance curves based on the quantum interference theory of graphene. We observe that the trigonal warping effect becomes weakened in the bulk as the Fermi energy of carriers moves towards the Dirac point. We observe also that the intravalley scattering in the confined cavity devices is enhanced, when the Fermi energy is in the close vicinity of the Dirac point, as a result of strong charge density fluctuations. The dephasing mechanism and the effect of edge scattering in the graphene cavity structures are also discussed.

\section*{Results and discussion}

Our devices were fabricated from high-quality single-crystalline monolayer graphene grown via chemical vapor deposition (CVD) \cite{LinLiRenEtAl2016}.
Subsequent to the growth, the graphene sheets were transferred onto highly doped silicon substrate capped with a 300-nm-thick SiO$_{2}$ layer, which was used as a
back gate. The transferred graphene devices exhibit a high electrical quality with the Hall mobility exceeding 25 000 cm$^2$V$^{-1}$s$^{-1}$ at T=1.9 K \cite{LinLiRenEtAl2016}. The multi-terminal Hall-bar and an inside cavity structure were patterned by means of electron beam lithography (EBL) followed by reactive
ion etching with oxygen plasma. Contact electrodes were subsequently fabricated by an additional EBL procedure and deposition of Ti/Au bilayer (10 nm/90 nm) by electron beam
evaporation. Figure 1(a) displays a false-colour atomic force microscopic (AFM) image of a device before depositing the electrodes. The dark yellow regions
correspond to trenches where graphene were etched out. The conductive regions of the graphene layer are shaded in light pink and green colours. Light blue parts are
remained graphene for depositing lateral side gate electrodes, which were not used in this study. The investigated device consists of two regions, one with cavity
and the other without cavity (bulk) for reference. The etched graphene cavity structure is highlighted by green colour in figure 1(a). Figure 1(b) shows The measurement configuration
used in this study. White lines are guides to the eyes for the boundaries of the etched graphene cavity and Hall bar structures. Bright
yellow strips are the metal electrodes. We applied constant current $I_{12}$ through electrodes 1 and 2, and simultaneously measured the voltage drop $V_{34}$
between probes 3 and 4 and the voltage drop $V_{56}$ between electrodes 5 and 6. This device geometry enables us to directly compare the transport properties of segments with
and without the cavity structure on the same device. The conductance across the cavity region is defined as $G_{34}=I_{12}/V_{34}$. We refer $G_{56}=I_{12}/V_{56}$
to the 'bulk' conductance, that is, for probing the transport properties of bulk graphene. When applying a small magnetic field with a negligible Hall resistivity, the definitions of $G_{34}$ and $G_{56}$ are still valid and accurate enough \cite{Datta1997}. The magnetotransport measurements were performed in a Physical Property
Measurement System (DynaCool) and a $^{3}$He/$^{4}$He dilution refrigerator, with a magnetic field, $B$, applied perpendicular to the plane of the graphene layer.
The four-terminal conductance was measured using a standard ac low-frequency lock-in technique with an excitation current of 10 - 100 nA at 13 Hz. Each cavity device consists of two 400-nm-wide constrictions and the diameter of the cavity is between 0.8 $\mu$m and 1 $\mu$m. We note that such wide constrictions are insufficient to act
as tunneling barrier and to confine a quantum
dot \cite{PonomarenkoSchedinKatsnelsonEtAl2008,SchnezMolitorStampferEtAl2009,GuettingerFreyStampferEtAl2010,WangCaoTuEtAl2010}. We observed Coulomb diamond
characteristics only when the cavity dimension is scaled down to 300 nm and the constriction width to 100 nm, as shown in figure S1. We have measured several cavity devices and
all the devices show similar features qualitatively. In this paper, we present the magnetotransport data obtained mainly from a device with a cavity diameter of 1
$\mu$m.

\begin{figure}[htbp]
\begin{center}
\includegraphics[width=6in]{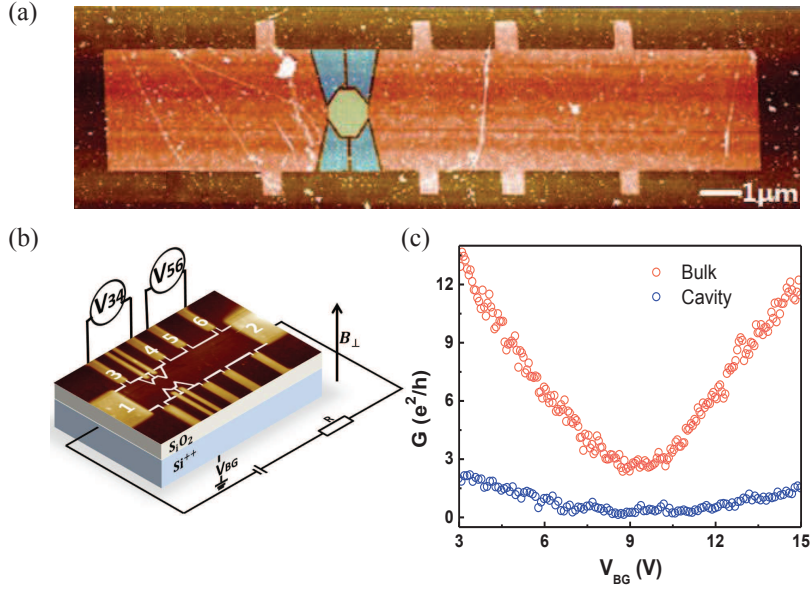}
\caption{(a) False-colour atomic force microscope image of a typical graphene cavity device. The graphene sheets were etched into a Hall bar with a width of 3 $\mu$m.  The etched cavity structure is highlighted by green colour in the image. The scale bar equals 1 $\mu$m.
(b) Schematic layout of our device and the measurement setup. Here, the region of the cavity and the boundaries of the Hall bar are outlined by the white lines. The current is driven from contact 1 to contact 2, while the voltages across the graphene cavity ($V_{34}$) and the bulk region ($V_{56}$) are measured simultaneously. (c) Four-point conductance as a function of the back gate voltage $V_{BG}$ measured at $T=$ 60 mK for the graphene cavity (blue circle) and bulk graphene (red circle).}
\label{fig1}
\end{center}
\end{figure}

Figure 1(c) displays the conductance of the cavity (blue circle) and the conductance of bulk graphene (red circle) in a typical device measured against the back gate voltage ($V_{BG}$)
at temperature $T=$ 60 mK. From the measured transfer characteristics and Hall measurements of the bulk graphene region of the device, we have determined the carrier density and plotted the conductance as
a function of carrier density, see figure S2. The graphene device is seen to be hole-doped with the Dirac point typically located at $V_{BG}\sim$ 9 V. The
transfer curves are not symmetric about the Dirac point, which is considered to be due to the influence of the misalignment at the electrode/channel interface and the
nonconstant density of states of the electrodes \cite{FarmerGolizadeh-MojaradPerebeinosEtAl2008}. Besides, charge transferring at the graphene/electrode interface
may also attribute to this asymmetry \cite{HuardStanderSulpizioEtAl2008}. In comparison to the bulk region, the conductance of the cavity region decreases
drastically and shows apparently a much broader transfer feature. This indicates that in the presence of the cavity structure, the carrier transport  is restricted to go through the constrictions and the cavity. Here, we note again that the confinement of the cavity is not strong enough to lead to quantized conductance or open a band gap
\cite{TerresChizhovaLibischEtAl2016,ChenLinRooksEtAl2007,HanOezyilmazZhangEtAl2007}.

Results of the magnetoconductance measurements performed at $T=$ 60 mK for both the cavity and the bulk regions under a perpendicular magnetic field are shown in figure 2. Figures 2(a) and 2(d) show the measured back gate voltage dependences of the conductance of the bulk and cavity regions, respectively. We denote some coloured dots at certain back gate voltages where we have made the magnetoconductance measurements. Figures 2(b) and 2(c) display the evolutions of the low field normalized magnetoconductance $\Delta G/G(0)=[G(B)-G(0)]/G(0)$, where $G(B)$ represents the conductance at magnetic field $B$ and $G(0)$ is the conductance at zero magnetic field, of the bulk region at different back gate voltages on the hole and the electron side (see the insets for the schematic band structure of graphene with different Fermi levels), while figures 2(e) and 2(f) show the corresponding results for the cavity region. In order to diminish the effect of universal conductance fluctuations, we have averaged the conductance over a range of 3.6 V in back gate voltage \cite{GorbachevTikhonenkoMayorovEtAl2007}. Note that we use the same average method to process the data in the following figures. The arrows beside the graphes indicate the direction of decrease in carrier density.

\begin{figure}[htbp]
\begin{center}
\includegraphics[width=6.0in]{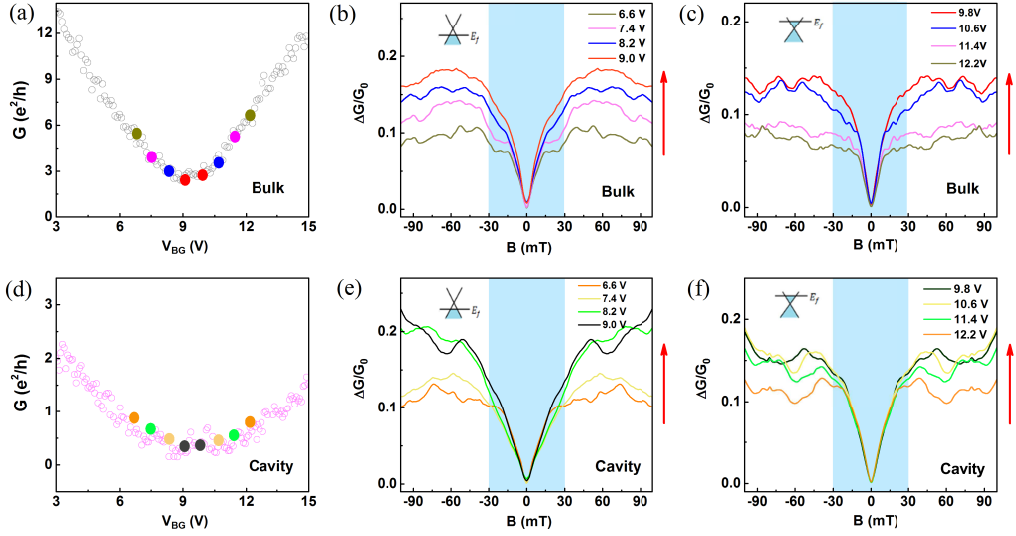}
\caption{Conductance of (a) the bulk and (d) the graphene cavity measured against the back gate voltage at $T=$ 60 mK and $B=$ 0 T. (b) and (c) Normalized magnetoconductance $\Delta G/G(0)=[G(B)-G(0)]/G(0)$ measured on the hole and electron transport sides for the bulk graphene at selected back gate voltages indicated by the coloured dots marked on the transfer curve in (a). (e) and (f) The same as (b) and (c) but  for the graphene cavity.
Every curve has been averaged over a back gate voltage window of $\Delta V_{BG}=3.6$ V for diminishing the effect of universal conductance fluctuations. The insets are schematics of $E_{F}$ positions set in the measurements.
The red arrows beside the graphes indicate the direction toward the Dirac point.}
\label{fig2}
\end{center}
\end{figure}

At comparatively lower magnetic field around B=0 T, a positive magnetoconductance with an obvious dip (see the shaded regions in figure 2) is clearly observed at all back gate voltages in both the bulk and the cavity regions, which can be ascribed to a WL effect. However, at higher magnetic field, the evolution of the magnetoconductance with varying carrier density exhibits different behaviors between the cavity and the bulk regions.
It can be seen that, for the bulk region, the curves gradually develop a downturn tendency with decreasing carrier density, indicating that WAL becomes strengthened as the back gate voltage approaches the Dirac point. This can be attributed to weakening the localization effect arising from trigonal warping in graphene \cite{AndoNakanishiSaito1998,McCannKechedzhiFal'koEtAl2006,SaitoDresselhausDresselhaus2000}, which breaks the $\overrightarrow{p}$ to $-\overrightarrow{p}$ symmetry in carrier dispersion and deforms equi-energy lines from circular contours to triangular contours within the same valley\cite{Fal'koKechedzhiMcCannEtAl2007,Peres2010},  around the Dirac point.
In the bulk, with decreasing density, the trigonal warping effect becomes weaker and thus results in enhancing WAL near the Dirac point. In contrast, the cavity region exhibits different evolution of the magnetoconductances with decreasing carrier density.
In figure 2(e), at $V_{BG}=6.6$ V and 7.4 V, the magnetoconductance curves become bending down with increasing magnetic field and exhibit the WAL feature.
When approaching the Dirac point, at $V_{BG}=8.2$ V and 9.0 V, the curves become upward bending indicating development of a tendency towards WL. The curves of the cavity region on the electron side show the same evolving trend, as seen in figure 2(f).


Based on both the measured transfer characteristics and the Hall measurements, we have extracted the mobility of our graphene samples at low temperature. The measured transfer characteristics give a mobility value of $\mu\approx11000$ cm$^2$V$^{-1}$s$^{-1}$ and a nearly same value is obtained from the Hall measurements, see the caption to figure S3. The corresponding mean free path extracted from the measurements is $l_{e} \approx 150-250$ nm, which is an order of magnitude smaller than
the cavity diameter (1 $\mu$m), such that the charge transport in the device is in the diffusive regime. These results allow us to analyze the measured data based on a 2D localization
theory. In order to get a more quantitative analysis, we fit our measured magnetotransport curves to the quantum interference
theory of monolayer graphene \cite{McCannKechedzhiFal'koEtAl2006}. The quantum correction to the conductance is given by
\begin{equation}
\Delta G(B)=\frac{e^2}{\pi h}\biggl[F\Bigl(\frac{\tau_B^{-1}}{\tau_\phi^{-1}}\Bigr)-F\Bigl(\frac{\tau_B^{-1}}{\tau_\phi^{-1}+2\tau_i^{-1}}\Bigr)-2F\Bigl(\frac{\tau_B^{-1}}{\tau_\phi^{-1}+\tau_i^{-1}+\tau_*^{-1}}\Bigr)\biggr],
\label{WL}
\end{equation}
where $F(z)=\ln(z)+\psi(0.5+z^{-1})$, $\psi(x)$ is the digamma function, $\tau_B^{-1}=4eDB/\hbar$ and $D$ is the diffusive coefficient.
Several scattering rates are considered, including inelastic scattering rate $\tau_\phi^{-1}$, elastic intervalley scattering rate $\tau_i^{-1}$ and $\tau_*^{-1}$. The rate $\tau_*^{-1}$ is related to the elastic intravalley scattering rate $\tau_s^{-1}$ and trigonal warping scattering rate $\tau_w^{-1}$ through the relation $\tau_*^{-1}=\tau_s^{-1}+\tau_w^{-1}$.
The first term with positive sign in equation ~(\ref{WL}) is responsible for weak localization, while the latter two terms with negative signs are responsible for weak antilocalization.
The solid lines in figures 3(a) and 3(c) are the best fits of the data to equation ~(\ref{WL}) with $\tau_\phi$, $\tau_i$ and $\tau_*$ as fitting parameters. The curves are successively  offset vertically for clarity. The magnetic field is restricted to the range of below 100 mT in the fittings in order to fulfill the small field requirement. It can be seen that our measured data is in good agreement with the theory.
In figures 3(b) and 3(d), the characteristic lengths $L_\phi$,\ $L_i$\ and $L_*$\ for the bulk and cavity regions, which are related to the fitted values of scattering rates via $L_{\phi,i,*}=(D\tau_{\phi,i,*})^{1/2}$, are plotted against the back gate voltage (the values of diffusive coefficient are plotted in figure S4). It is clearly seen that $L_\phi$\ in both the bulk and cavity regions become shorter as the carrier density decreases, suggesting enhanced dephasing of carriers near the Dirac point. As approaching the Dirac point, the formation of electron-holes puddle in graphene leads to inhomogeneities modifying the geometry of conducting paths and destroys the constructive interference of charge carriers \cite{TikhonenkoHorsellGorbachevEtAl2008,ZhangBrarGiritEtAl2009}. Besides, the puddles can also introduce fluctuating electromagnetic fields,  which enhance the Nyquist scattering, leading to the destruction of phase coherence \cite{KiJeongChoiEtAl2008}.

\begin{figure}[htbp]
\begin{center}
\includegraphics[width=6.0in]{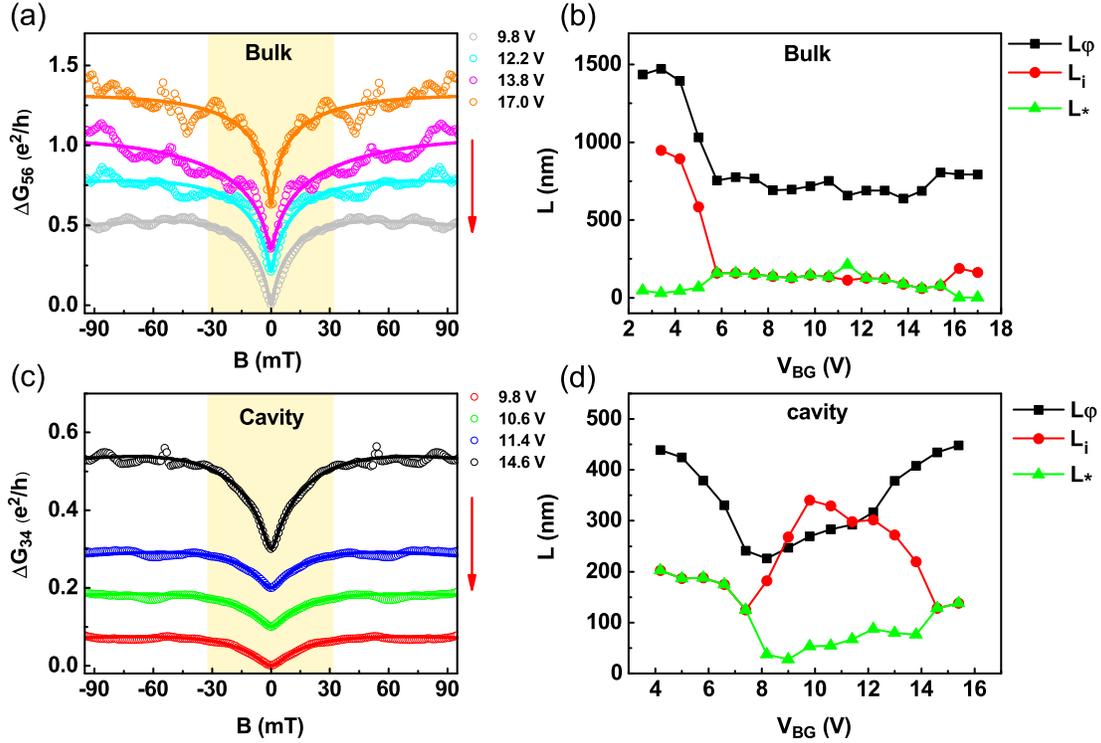}
\caption{(a) and (c) Magnetoconductance measured at temperature of 60 mK at various applied back gate voltages for the bulk and the cavity regions. The arrow denotes the direction toward the Dirac point.
The curves are vertically shifted for clarity.
Solid lines are the best fits of the magnetoconductance curves to equation ~(\ref{WL}).
(b) and (d) The characteristic lengths $L_\phi$, $L_i$ and $L_*$ as a function of back gate voltage in the bulk and the cavity. It can be clearly seen that $L_i$ and $L_*$ in the bulk and the cavity show very different evolutions with back gate voltage near the Dirac point.
}
\label{fig3}
\end{center}
\end{figure}

A remarkable feature of the data presented in figures 3(b) and 3(d) is that $L_i$ and $L_*$ for the bulk and cavity regions show different evolution trends as a function of back gate voltage.
For the bulk region, it is seen that $L_i$ ($L_*$) decreases (increases) with shifting back gate voltage towards to the Dirac point.
However, for the cavity region, $L_i$ and $L_*$ exhibit opposite evolution trends, suggesting that elastic scattering mechanisms near the Dirac point are different in the bulk and cavity regions.


Further insight into the role of elastic scattering mechanisms on magnetotransport can be gained by plotting the ratio of intra- and intervalley scattering rates.
In the theory of McCann \emph{et al.}, the magnetoresistance behavior in graphene is shown to depend on the ratio of $B_{*}$ and $B_{i}$, where $B_*=\frac{\hbar}{4eD\tau_{*}}$ and $B_{i}=\frac{\hbar}{4eD\tau_{i}}$ are the characteristic transport fields related to intravalley and intervalley scattering, respectively \cite{McCannKechedzhiFal'koEtAl2006}.
Recently, Tikhonenko \emph{et al.} experimentally determined a diagram \cite{TikhonenkoKozikovSavchenkoEtAl2009} of the scattering times related to the transition between WL and WAL in agreement with the theory \cite{McCannKechedzhiFal'koEtAl2006}. In figures 4(a) and 4(b) the ratio of $B_{*}/B_{i}$ is plotted as a function of the back gate voltage for the bulk and cavity regions, respectively.
As shown in figure 4(a), for bulk region, one can see that in the vicinity of the Dirac point, the $B_*/B_i$ value is about 1, that is, $B_*$ and $B_i$ have almost the same values, which is consistent with $L_*\approx L_i$ in figure 3(b) and previous measurements \cite{ChenBaeChialvoEtAl2010}.
Away from the Dirac point, $B_*/B_i$ becomes larger, implying that intravalley scattering may be stronger than intervalley scattering at high carrier densities. On the whole, the curve of ratio $B_*/B_i$ exhibits a '$\bigcup$' shape.
The intravalley scattering term $B_*$ is decided by  two rates $\tau_w^{-1}$ and $\tau_s^{-1}$ since $\tau_*^{-1}=\tau_s^{-1}+\tau_w^{-1}$. We attribute the increasing $B_*$ values to the strengthened trigonal warping effect caused scattering rate $\tau_w^{-1}$ as the Fermi level goes away from the Dirac point.

\begin{figure}[htbp]
\begin{center}
\includegraphics[width=6.0in]{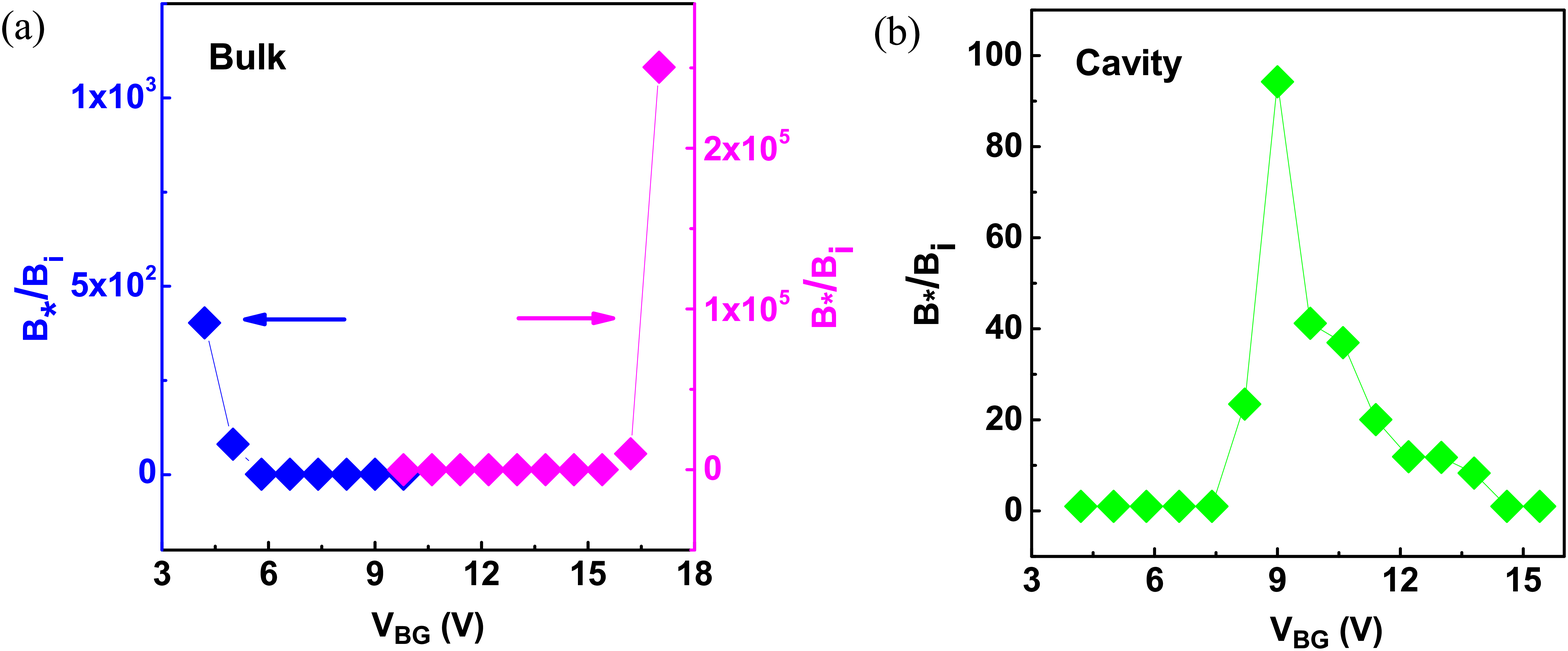}
\caption{(a) and (b) Ratios of $B_*/B_i$ as a function of the back gate voltage extracted for the bulk and the cavity region.
$B_*$ and $B_i$ are the characteristic transport fields relating to intravalley and intervalley scattering, respectively. Blue (purple) rhombuses represent the ratios on the hole (electron) side with their values given in the left (right) axis in (a).
In contrast to the bulk, $B_*/B_i$ for the graphene cavity increases steeply when approaching the Dirac point, indicating an enhancement of the intravalley scattering. The lines in (a) and (b) are guides to the eyes.
}
\label{fig4}
\end{center}
\end{figure}

Compared to the bulk region, $B_*/B_i$ for the cavity region increases steeply when approaching the Dirac point and the curve shows a '$\bigcap$' shape.
This implies that intravalley scattering becomes dominant as the Fermi level gets close to the Dirac point which is completely contrary to what happens in the bulk.
In graphene, $\tau_w^{-1}$ becomes smaller when Fermi level shifts to the Dirac point leading to a suppression of $B_*$. In contrast, the observation of enhanced $B_*$ near the Dirac point for the cavity region indicates the existence of other mechanisms related to enhanced intravalley scattering rate.

Here, we propose that the enhanced intravalley scattering is associated with the effect of charge inhomogeneity and edge disorder on the confined graphene nanostructures. We point out that charge inhomogeneity could be mainly induced by structural distortions and chemical doping from
fabrication residues \cite{ObraztsovObraztsovaTyurninaEtAl2007,ChaeGuenesKimEtAl2009,CaladoSchneiderTheulingsEtAl2012,Schwierz2010,XueSanchez-YamagishiBulmashEtAl2011}, which have the length scale being comparable to the dimension of the cavity. On the other hand, recent experimental studies
have shown that graphene edges have an important influence on transport in reactive ion etched graphene
nanodevices \cite{BischoffLibischBurgdoerferEtAl2014,BischoffSimonetVarletEtAl2016}. The transport measurements of graphene nanostructures have revealed that electrons can localized along the edges of etched graphene on length scales much longer than the physical disorder length \cite{BischoffLibischBurgdoerferEtAl2014} and comparable to the dimension of our cavity device. Besides, zigzag type edges allowing for intravalley scattering by changing psesudospin might also have contributed to intravalley scattering \cite{ParkYangMayneEtAl2011}. As the Fermi level shifts close to the Dirac point, it has two main impacts on the intravalley scattering in the cavity region. First, due to the low density of states around the Dirac point, the screening of the charged impurity scattering centers becomes weakened. Thus, carriers see more scattering centers and, hence, intravalley scattering rate $\tau_s^{-1}$ is enhanced, leading to larger $B_*$. Second, the cavity is a comparatively closed confined system with a narrow entrance and exit formed by the two constrictions. Carriers are likely scattered back and forth by the edges of the cavity and experience a long path and dwelling time before getting out, and hence gain more chances being scattered by the charge inhomogeneity and edge disorders, resulting in enhanced intravalley scattering in the cavity region. 
To exclude the influence of the constriction and confirm the confined effect of the cavity structure on the observed weak localization features, we have also performed the same magnetotransport measurements on a single constriction graphene device, which are shown in figure S5 in the supporting information.
In previous scanning tunnelling microscopy (STM) measurements, Zhang \emph{et al.} have probed the intravalley scattering processes near the charged puddles \cite{ZhangBrarGiritEtAl2009}. They revealed that the charge puddles can also act as intravalley scattering centers. However, it is also worth noting that the characteristic length scale of such charge puddles is $\sim$20 nm \cite{MartinAkermanUlbrichtEtAl2008,DeshpandeBaoZhaoEtAl2011,ZhangBrarGiritEtAl2009} and can not contribute to the observed transport difference between the cavity and bulk graphene. Further experimental and theoretical studies are needed to understand the contributions from intrinsic charge puddles to the intravalley scattering in graphene nanostructures.


So far we have presented that the elastic intravalley scattering process is enhanced in a confined graphene cavity device.
To study the confinement effect on the inelastic scattering, we measured the temperature dependent magnetoconductance in the cavity and bulk regions. In figure 5(a), the magnetoconductance of the cavity region is plotted for various temperatures ranging from 2.5 to 27 K. We have applied the same analysis and fitted the magnetoconductance data in the framework of the quantum interference theory of graphene. The best fits of the magnetoconductance traces of the cavity to equation~(\ref{WL}) are shown as solid lines in figure 5(a). Figures 5(b) and 5(c) show the extracted dephasing rate $\tau_\phi^{-1}$ as a function of temperature for the bulk and cavity regions, respectively. As we can see, the dephasing rate of the cavity region shows different temperature dependence from the bulk region, suggesting that they arise from different dephasing mechanisms. Previous studies have shown that the main dephasing mechanism in graphene at low temperatures is electron-electron interaction \cite{GonzalezGuineaVozmediano1996,GonzalezGuineaVozmediano1999,SarmaHwangTse2007,PoliniAsgariBarlasEtAl2007}.
The inelastic scattering by the electron-electron interaction can be divided into two terms. One is the direct Coulomb interaction among the charge carriers \cite{TaboryskiLindelof1990}. The other is Nyquist scattering, which origins from the electrons interacting with the fluctuating electromagnetic field induced by the random movement of neighboring electrons \cite{TaboryskiLindelof1990,HansenTaboryskiLindelof1993}.
The expression \cite{KiJeongChoiEtAl2008} for the the dephasing rate is given by
\begin{equation}
\begin{gathered}
1/\tau_{\phi}=1/\tau_{N}+1/\tau_{ee}+const=ak_{B}T\frac{\ln(g)}{\hbar g}+b\frac{\sqrt{\pi}}{2\nu_{F}}(\frac{k_{B}T}{\hbar})^2\frac{\ln(g)}{\sqrt{n}}+const,
\label{Time}
\end{gathered}
\end{equation}
where $\tau_N^{-1}$ is the Nyquist scattering rate, $\tau_{ee}^{-1}$ is the direct Coulomb interaction rate which corresponds to large moment transfer collision, the coefficient $a$ and $b$ are dimensionless parameters and independent of temperature which represent the strength of these two kinds of scattering, $\nu_{F}$ is the Fermi velocity with a value $\nu_{F}\simeq1\times10^6$ m/s, $k_{B}$ is the Boltzmann constant, $g(n)$ is the normalized conductivity defined as $g(n)=\sigma(n)h/e^2$. The red dashed line in figure 5(b) is the linear fit of $\tau_\phi^{-1}$, suggesting that the Nyquist scattering is the dominant source of dephasing for bulk region at low temperatures. Previous experimental studies on graphene have also demonstrated a similar behavior \cite{TikhonenkoHorsellGorbachevEtAl2008,KimShinParkEtAl2016}. In contrast, as clearly shown in figure 5(c), $\tau_\phi^{-1}$ follows a parabolic temperature dependence in the cavity region, indicating that the direct Coulomb interaction scattering mechanism governs dephasing in a graphene cavity device.

\begin{figure}[htbp]
\begin{center}
\includegraphics[width=6.0in]{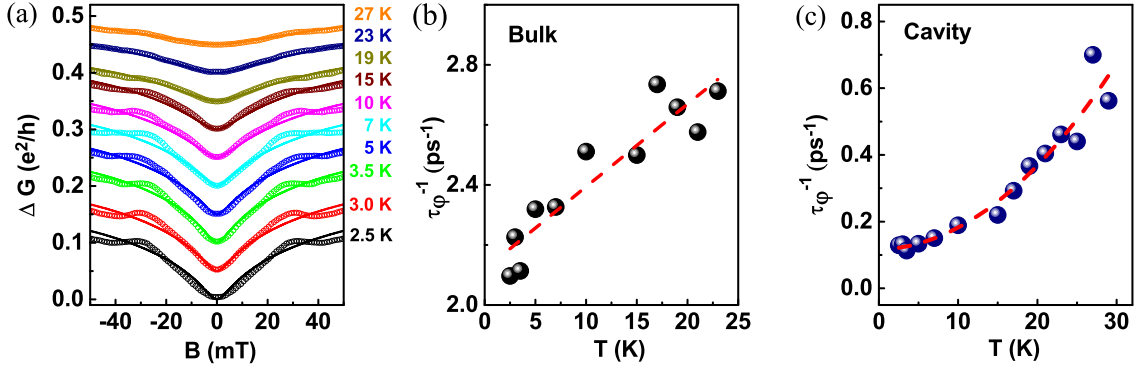}
\caption{(a) Magnetoconductance of the cavity device measured at $V_{BG}=2$ V and at different temperatures. The curves are successively offset vertically for clarity.
(b) and (c) Temperature dependence of ${\tau_\phi^{-1}}$  extracted from fitting the measured data to equation~(\ref{WL}) for the bulk and graphene cavity.
The red dotted line in (b) represents a linear fit of ${\tau_\phi^{-1}}$ for the bulk region.
In contrast, ${\tau_\phi^{-1}}$ for the graphene cavity follows a parabolic temperature dependence, see the fitted red dotted line in (c), suggesting that the direct Coulomb interaction scattering mechanism governs
the dephasing at low temperatures.}
\label{fig5}
\end{center}
\end{figure}

The presence of the strong direct Coulomb interaction is expected to result in a large momentum transfer process of inelastic scattering \cite{KiJeongChoiEtAl2008}. For the graphene cavity, we expect that the edges of confined graphene cavity structures are the likely source of the enhanced inelastic scattering process. This is also supported by previously reported Raman spectroscopy results \cite{CancadoPimentaNevesEtAl2004,CasiraghiHartschuhQianEtAl2009a}.
On etched graphene nanoribbons, strong D peak in the Raman spectra is observed in the edge region which is known due to the double resonance process activated by inelastic intervalley scattering \cite{ThomsenReich2000,SaitoJorioSouzaFilhoEtAl2001,BischoffGuettingerDroescherEtAl2011}.
For our oxygen plasma etched graphene cavity, there are two main impacts on the inelastic scattering process. One is that carriers gain increased probability to be scattered by the edges because of the small dimension of the cavity structure compared to a open system. The other is that carriers get the chance to be scattered many more times with the edges inside the cavity before getting out because of its confined structure. It is therefore likely that the inelastic intervalley scattering is enhanced by the cavity edges, leading to the observed $T^2$ dependency of $\tau_{\phi}^{-1}$.

\section*{Conclusions}
In summary, we have performed low-temperature magnetotransport measurements of graphene cavity devices made from high-quality graphene grown by CVD. We find that
the confined cavity structures have prominent effects on the magnetoconductance of the devices. The observed intravalley scattering
enhancement near the Dirac point in cavity devices can be attributed to the effect of charge inhomogeneities and edge disorder in the confined graphene
nanostructures. We also show that graphene edges play an important role in the inelastic dephasing process in the cavity devices. Our results reveal the
importance of confinement in the electrical transport of graphene nanostructures, which offer opportunities for realizing graphene-based nanoelectronics and for investigating the transport properties of confined Dirac fermions in graphene.

\section*{Acknowledgements}
We acknowledge financial supports by the Ministry of Science and Technology of China (MOST) through the National Key Research and Development Program
of China (No. 2016YFA0300601 and 2017YFA0303304), and National Natural Science Foundation of China (Nos. 91221202, 11374019, 91421303, and 11774005). \\
\section*{References}
\bibliography{iopart-num}

\end{document}